\documentclass[
 reprint,
 amsmath,amssymb,
 aps,
]{revtex4-1}

\usepackage{graphicx}
\usepackage{dcolumn}
\usepackage{bm}

\usepackage{caption}
\usepackage{color}

\def\mbf#1{\mbox{\boldmath ${#1}$}}

\begin{document}

\preprint{APS/123-QED}

\title[Identifying information in DNN]{Identifying Cosmological Information in a Deep Neural Network}

\author{Koya Murakami}
 \email{murakami.koya@a.mbox.nagoya-u.ac.jp}
 \affiliation{Department of Physics, Nagoya University, Furocho, Chikusa, Nagoya, 464-8602, Japan}
 
\author{Atsushi J. Nishizawa}
 \email{atsushi.nishizawa@iar.nagoya-u.ac.jp}
\affiliation{
 Department of Physics, Nagoya University, Furocho, Chikusa, Nagoya, 464-8602, Japan,\\
 Institute for Advanced Research, Nagoya University, Furocho, Chikusa, Nagoya, 464-8602, Japan}

\date{\today}

\begin{abstract}
A novel method images to estimate cosmological parameters based on images is presented. In this paper, we demonstrate the use of a convolutional neural network (CNN)
for constraining the mass of dark matter particle. For this purpose, we perform a suite of N-body simulations with different 
dark matter particle masses
to train CNN and estimate 
dark matter mass using a density-contrast map. The proposed method is complementary to the one based on 
summary statistics, such as two-point correlation function. We compare our CNN classification results
with those obtained from
the two-point correlation of the distribution of
dark matter particles, and find that the CNN offers better performance
In addition, we use images made from a rondom Gauss simulation 
to train a CNN, which  
is then compared with the CNN trained by N-body simulation and two-point correlation.
The random Gauss-trained CNN has comparable performance to two-point correlation.
\end{abstract}

\pacs{Valid PACS appear here}
\maketitle

\section{\label{sec:intro}Introduction}
At present, $\Lambda$CDM model is widely accepted in cosmology. 
This model assumes that dark matter (DM) is cold, i.e., its mass is heavy enough that DM particles were non-relativistic at the time freeze-out, however, it dose not make any concrete assumptions about the mass of DM.
This mass is an essential parameter for determining the correct model of DM. For example, in sterile neutrino DM models, the mass ranges from 1 keV to 1 MeV  \cite{Boyarsky2019}; in weakly interacting massive particle (WIMP) model, the mass ranges from 10 GeV to 1 TeV \cite{Alvarez2020}.
Because the DM mass affects the large-scale structure of  the universe at small scale,
it is constrained by the Lyman-$\alpha$ forest power spectrum, which shows that its mass must be heavier than $\mathcal{O}(1)\  \mathrm{keV}$ \cite{Garzilli2019,Garzilli2019a}. However, this constraint is insufficient to choose between DM models, so we need a new method capable of gaining more information about large-scale structure.

In this paper,
we focus on constraining on the mass of DM using a neural network (NN). NNs are machine-learning (ML) algorithms used for big-data analysis.
NNs learn how to extract information from much labeled-data without humans deciding which features of the data to use. There are various kinds of NNs. Convolutional neural networks (CNN) are used to extract information from images using filters.
For example, CNNs are used to distinguish the images of a dog from those of a cat and to detect human faces in an image with extremely high accuracy.

NNs are also used in cosmology. Typical analytical techniques, such as two-point correlation of the matter-density distribution, can only obtain part of the information from observed data; however, an ML algorithm can extract complex information from the data and capture various important  features. For example, a CNN has been used to constrain cosmological parameters in the fields of weak lensing cosmology \cite{Ribli2019a}, simulated convergence maps \cite{Ribli2019}, and  large-scale structure \cite{Pan2019}. As other examples, the signals of the Sunyaev-Zel'dovich effect are detected with U-Net, a network which first extracts the feature and then applies up-convolution to retain the original image resolution \cite{Bonjean2020}, modified gravity models are distinguished from the standard model using  CNNs \cite{Peel2019}, and the initial conditions of the universe are reconstructed by NNs using galaxy positions and luminosity data \cite{Modi2018}.
NNs used in these previous works have shown better performance than typical analysis.

This paper provide a potential that CNN can constrain on the mass of dark matter particle more strongly. The Lyman-$\alpha$ power spectrum only focus on two-point correlation of the structure at the small scale. On the other hand, CNN can extract additional information from images of the large scale structure. Therefore, we study the performance of classifying images from simulations by CNN.

This paper is organized as follows: In Section \ref{sec:simulation}, we introduce non-cold DM  models, describe our simulation suite, and construct the training and validation dataset; in Section \ref{sec:method}, we show the calculation of two-point correlation and our CNN architecture and the calculation method performed by our CNN; and in Sections \ref{sec:result}, \ref{sec:discussion}, and \ref{sec:summary}, we discuss the results of our CNN analysis and summarize our work.

Throughout this paper, we use the cosmological parameters taken from Planck 2018 \cite{Akrami2018}, except for the mass of the DM particle.

\section{\label{sec:simulation}Simulations and Initial Conditions}
\subsection{\label{ssec:wdm}Non-cold Dark Matter Model}
\begin{figure}
    \centering
    \includegraphics[width=\linewidth]{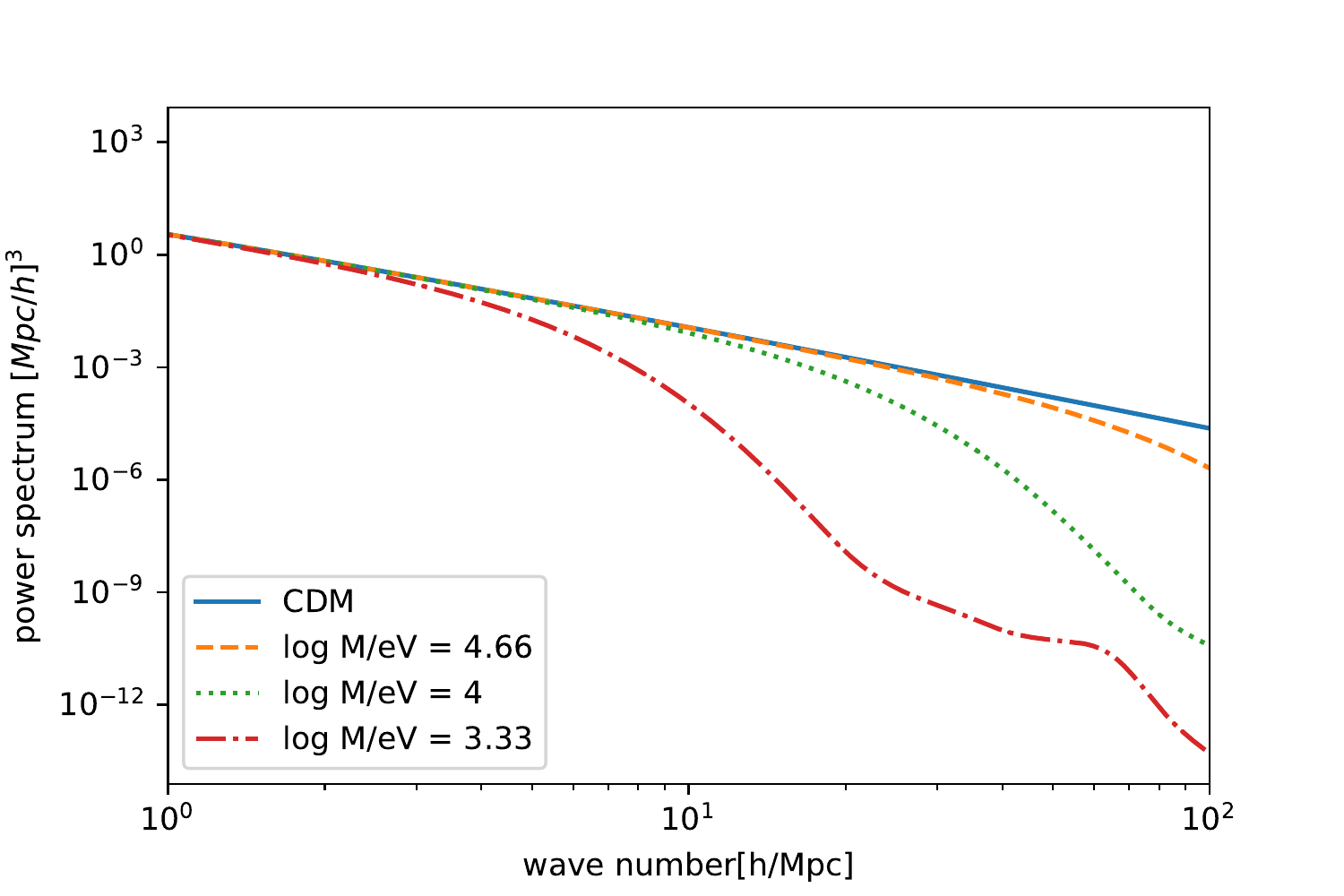}
    \caption{Examples of the input matter power spectra for the simulations. The lighter the non-cold dark matter is, the lower the amplitude of the small-scale power spectrum is.
    }
    \label{fig:ini_powspec}
\end{figure}

DM has a non-zero mass, but does not interact with electromagnetic radiation (or at least interacts with it very weakly). Therefore, we have not observed DM directly and it can only be seen by its gravitational interactions. The formation structure in the universe is affected by DM's gravity, so we obtain information about DM by observing the large-scale structure of the universe.

In this paper, we consider cold dark matter(CDM) and non-cold dark matter(NCDM). 
CDM is heavy enough to be a non-relativistic particle at the time of freeze-out of DM particle; its velocity dispersion is negligible. 
On the other hand, NCDM is a light particle, which experiences significant velocity dispersion.

DM velocity dispersion prevents the structure of the mass-density distribution from growing, especially at the small scale. The velocity dispersion is $\propto 1 / m_\chi$ ($m_\chi$ is DM mass), and the dumping scale of the matter power spectrum caused by this velocity dispersion is $\propto m_\chi$ \cite{Boyanovsky2011} (See Fig. \ref{fig:ini_powspec}).

We calculate the matter power spectrum using Cosmic Linear Anisotropy Solving System (\texttt{CLASS})  \cite{Lesgourgues2011a} for various DM models. 
In \texttt{CLASS}, the density perturbation is calculated based on Ma and Bertschinger(1995)  \cite{Ma1995}. The calculation for the CDM model is simple, as CDM is treated as a pressureless perfect fluid. 
The NCDM models, however, are more complicated. In \texttt{CLASS}, NCDM is treated as a sterile neutrino, which is a fundamental particle added to the standard model and is distinguished from active neutrinos (electron, mu, and tau neutrino). \texttt{CLASS} rescales the Fermi-Dirac distribution based on the widely studied sterile neutrino model  \cite{Dodelson1994} and calculates the time evolutions of density perturbations $\delta$, fluid-velocity divergences $\theta$, and shear stress $\sigma$ in the phase space using the fluid approximation \cite
{Lesgourgues2011b} (see Section 3 in \cite{Lesgourgues2011b}).

In this work, we consider the CDM model and nine NCDM models with different DM masses.

\subsection{\label{ssec:sim}Implementation to N-body Simulation}
We perform a set of N-body simulations for DM models with different particle masses.
We assume that in our DM model, after redshift $z=20$, where the initial condition for our N-body simulation is generated, all the particles interact only with gravitational force. All features of DM models can be encoded into the matter power spectrum at the initial condition. We set the cosmological parameters obtained by Planck \cite{Akrami2018} as $\Omega_m=0.311,\ \Omega_\Lambda=0.689,\ \Omega_b=0.049,\  h=0.677,$ and $\ln 10^{10} A_s=3.047$. In addition to the standard CDM model, we perform simulations with NCDM (non-CDM) models with $m_{\chi}$ being logarithmically uniformly sampled from $10^{2.33}$ to $10^5$ [eV].

The matter power spectrum for the initial condition of the N-body simulation is calculated by \texttt{CLASS} \citep{Lesgourgues2011a}, as shown in Fig. \ref{fig:ini_powspec}. With these input power spectra, we generate the initial condition using \texttt{2LPTic} \citep{Crocce2006} and apply glass realization to erase the gridding pattern in the particle distribution. To solve the gravitational evolution, we use \texttt{Gadget-2} \citep{Springel2005} with a box size of 200 Mpc/$h$ on a side, for $1024^3$ particles. The initial condition of the simulation is generated at $z=20$ and the simulation is terminated at $z=0.3$.
We conduct two independent N-body simulation run with different random seeds for the initial condition.

We use one realization for training, and the other as a totally independent test dataset to evaluate our method.

\subsection{\label{ssec:train}Training and Test Sets}

\begin{figure*}
  \centering
  \includegraphics[keepaspectratio,width=\linewidth]{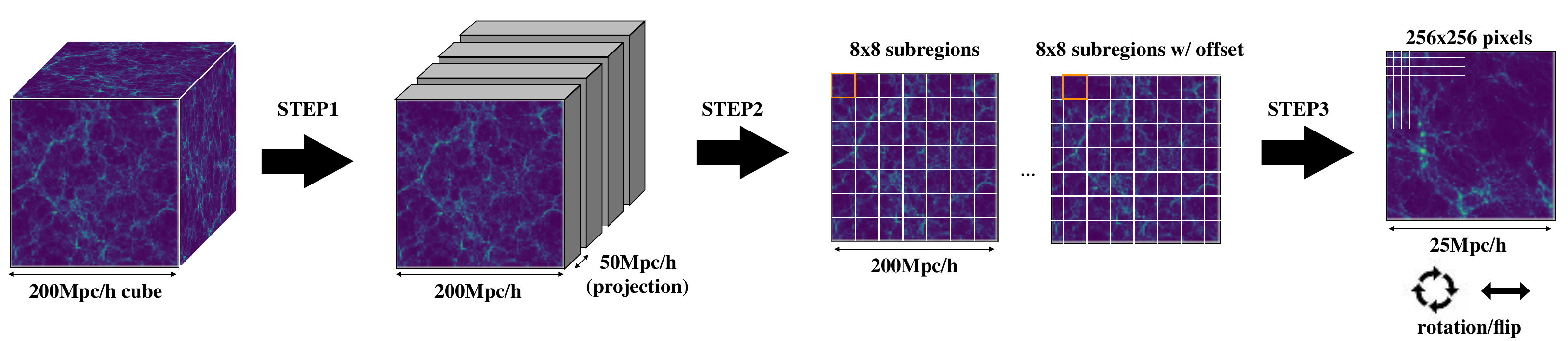}
  \caption{The workflow for making images from simulation data
  }
  \label{fig:gen_img}
\end{figure*}

\begin{figure}
    \centering
    \includegraphics[keepaspectratio,width=\linewidth]{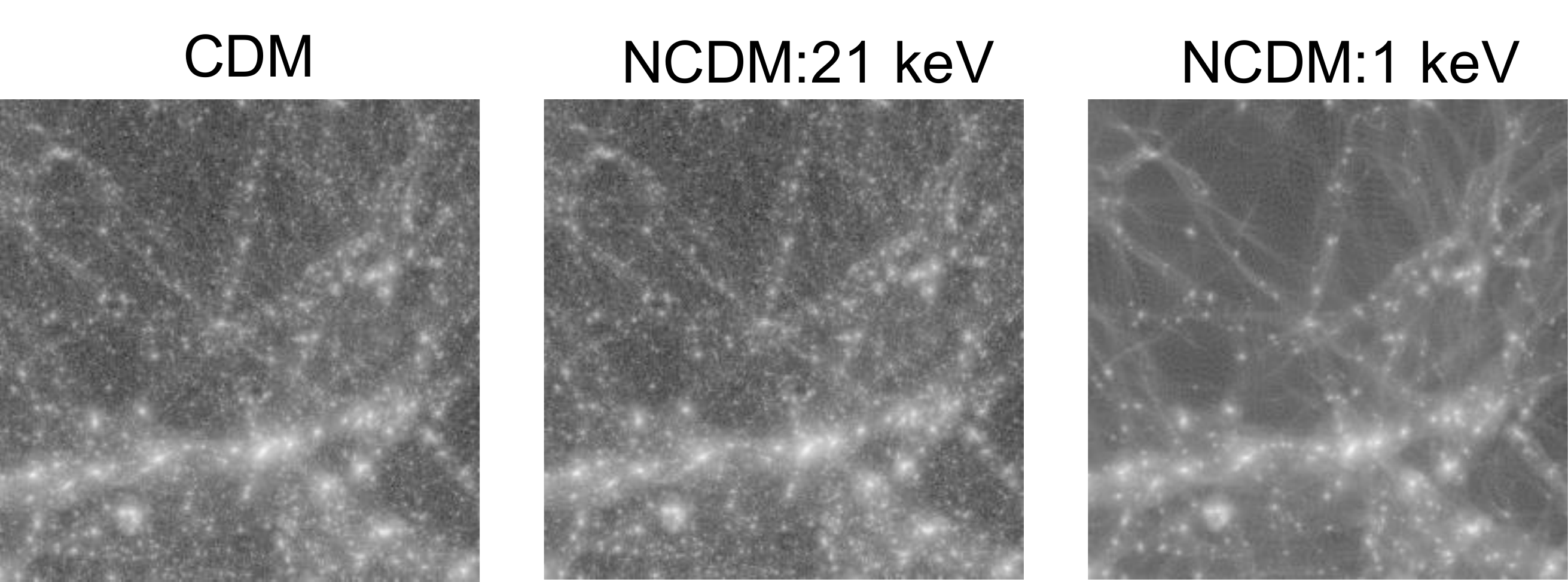}
    \caption{Example images of the CDM, 21 keV NCDM, and 1 keV NCDM models. These images are made from the same region of the simulation box.}
    \label{fig:image}
\end{figure}

In this subsection, we describe the procedures for constructing the images from N-body simulation used for training and validation. The processes for constructing the test sample are completely the same except that the dataset is independent realization of the simulation.
The damping scale due to the free-streaming of NCDM for $10^{2.33}$ eV is about $k_{fs}\sim 1h/$Mpc and $k_{fs}\sim 100h/$Mpc for $10^5$ eV DM. Therefore, the image size should be sufficiently large to include the mode $k\sim 1\ h/$Mpc, and at the same time it should have sufficient resolution to resolve $k\sim 100 h/$Mpc mode fluctuations. Our choice of the box size and number of particles suffices for these requirements.
Here we assume practical observation where the distance to the objects is not well accurately determined like in the case of  multi-band imaging observation.

To generate images from N-body simulation, we implement the following procedures (see Fig. \ref{fig:gen_img}):
\begin{enumerate}
  \item We divide the simulation box into four pieces along the line of sight with each width being $50$ Mpc/$h$. Within each sub-region, the particle positions along the line of sight are projected onto the $x$-$y$ plane, approximating the situation of inaccurate determination of galaxies' redshift. This uncertainty roughly corresponds to $\Delta z=0.01$ at $z=1$, which is going to be achieved by photometric redshift of future imaging surveys such as Euclid \citep{Euclid} or Rubin Observatory LSST \citep{LSST}. We have three degrees of freedom for choosing the line of sight direction; these can be considered independent realizations.
  \item For each slice of the simulation, we further divide the region into $8\times8$ small patches, with each patch including the typical structure of the large-scale structure; i.e., one patch has 25 Mpc/$h$ on a side. To increase  sampling, we employ multiple offsets when we subdivide the slices. The offsets are
  $\Delta = 25i/256$ Mpc/$h$ where $i=1,2,\cdots,50$ in the $x$ or $y$ directions.
  This may increase the number of available images sufficiently and significantly help our training process to converge,
  although the shifted images are not totally independent of each other.
  \item Finally, we rasterize the particle data into one image. In each patch, we assign particles to $256\times 256$ pixels using the nearest gridding point (NGP) to obtain a  $\rho(\boldsymbol{x})$ map. Then, $\rho(\boldsymbol{x})$s in the map are converted to $\delta(\boldsymbol{x}) = (\rho(\boldsymbol{x}) - \bar{\rho}) / \bar{\rho}$ where $\bar{\rho}$ is the mean $\rho(\boldsymbol{x})$ over the simulation box.
\end{enumerate}
In total, we have (3 line of sight directions) $\times$ (4 redshift slices) $\times$ ($8^2$ patches) $\times$
($100$ offsets) = 76,800 images for one realization of the N-body simulation. In addition, in training the CNN, 
the images are rotated every 90 degrees and flipped to generate another different set of images.
Thus, the number of training dataset is effectively $76,800$ $\times$ ($2$ flips) $\times$ ($4$ rotation) $=614,400$; however, in testing our CNN, test images are not flipped or rotation.
The image of density fluctuation $\delta$ has large dynamic ranges due to the non-linear evolution of the structure. For our neural network architecture, it is difficult to extract feature quantities from such high dynamic range images; therefore, we apply the transformation
\begin{equation}
    \delta'(x) = {\rm arcsinh} [\delta(x)].
\end{equation}
This transformation is motivated by the magnitude system, \textit{Luptitude} introduced by the Sloan Digital Sky Survey \citep{Lupton+:1999}. This is particularly useful for reducing the dynamic range, including negative values to which a simple logarithmic scale cannot be applied.

\subsection{\label{ssec:simRG}Random Gaussian Simulations}
This paper mainly aims to identify the source of information extracted through a deep neural network, in addition to the power spectrum. To this end, we prepare data that exactly obey the same power spectrum as the N-body simulations with a density realization given by a totally random Gaussian field.

First we measure the 2D power spectrum $P(k_\perp)$ from our N-body simulation snapshots, projected along the line of sight with 50 Mpc/$h$ width; then we generate the random variable $\tilde{\delta}(\boldsymbol{k}_\perp)$, which obeys a Gaussian distribution with
$\langle\tilde{\delta}\rangle=0$, and $\langle\tilde{\delta}^2\rangle=P(k_\perp)$. The gridding in $k-$space is $\Delta k$=0.25$h/$Mpc, so we reproduce the same image resolution with the N-body simulation, $25/256\sim 0.1 {\rm Mpc}/h$. Using the same processes, we generate 76,800 $\times$ $2$ (flips) $\times$ $4$ (rotation) multiple images from the single realization of random Gaussian simulation. Test images are generated in exactly the same manner, except that a different random seed is used.

\section{\label{sec:method}Method}
\subsection{\label{ssec:xi}Two-Point Correlation}
A large amount of cosmological analysis has been done mainly using two-point statistics, such as the power spectrum of two-point correlation functions. In this paper, we examine how much information can be extracted from the raw image data in comparison to summary statistics of this sort. To compare our results from the convolutional neural network, we first introduce the projected two-point correlation function,
\begin{equation}
 w_{\rm P}(r_{\rm P}) \equiv \langle \delta(\mbf{r_1}) \delta(\mbf{r_2})\rangle ,
\end{equation}
where $\delta$ is the density fluctuation of matter, $r_{\rm P}=|\boldsymbol{r}_1-\boldsymbol{r}_2|\cos\theta_{12}$ is the projected separation, and square parentheses denote the ensemble average over the different realizations of the universe, here we replace this ensemble average with the spatial average over the entire simulation box.

To keep the available input information consistent with what we will use in the deep neural network, we compute the two-dimensional projected correlation function based on the projected particle position. $w_{\rm P}$ is measured using the publicly available code \texttt{TreeCorr} \cite{2015ascl.soft08007J}. The projected separation spans from the image size to the image resolution, namely, 25 Mpc/$h$ to 25/Ngrid (Ngrid$=256$) Mpc/$h$, which is equally separated into 32 bins along the logarithmic scale. For the covariance measurement, we use jackknife resampling, with each sub-region corresponding to one image, $25\times 25 [{\rm Mpc}/h]^2$ in area. Thus, we have $8^2$ (images) $\times$ $4$ (line of sight slices) = $256$ subsamples. If we use  $w_{{\rm P},k}^{\rm JK}$ to denote the $k$-th jackknife measurements, the covariance matrix can be written as
\begin{equation}
  \label{eq:jk_cov}
  \begin{split}
    \mathrm{C}_{ij} = \frac{n-1}{n} \displaystyle\sum^{n}_{k}
    [ & w_{{\rm P},k}^{\rm JK}(r_{i})- \overline{w_{\rm P}^{\rm JK}}(r_{i})]  \\
    &\times [w_{{\rm P},k}^{\rm JK}(r_{j})-\overline{w_{\rm P}^{\rm JK}}(r_{j})] \ ,
  \end{split}
\end{equation}
where $\overline{w_{\rm P}^{\rm JK}}$ is the mean $w_{\rm P}$ over jackknife resampling.

\subsection{\label{ssec:dnn}Convolutional Neural Network}
\begin{table}[h]
  \begin{tabular}{| c || c | c |}
    \hline
    \ & Layer & Output map size \\ \hline
    1 & Input & $256 \times 256 \times 1$ \\
    2 & $3 \times 3$ convolution & $254 \times 254 \times 32$ \\
    3 & $3 \times 3$ convolution & $252 \times 252 \times 32$ \\
    4 & $3 \times 3$ convolution & $250 \times 250 \times 64$ \\
    5 & $3 \times 3$ convolution & $248 \times 248 \times 64$ \\
    6 & $2 \times 2$ AveragePooling & $124 \times 124 \times 64$ \\
    7 & $3 \times 3$ convolution & $122 \times 122 \times 128$ \\
    8 & $1 \times 1$ convolution & $122 \times 122 \times 64$ \\
    9 & $3 \times 3$ convolution & $120 \times 120 \times 128$ \\
    10 & $2 \times 2$ AveragePooling & $60 \times 60 \times 128$ \\
    11 & $3 \times 3$ convolution & $58 \times 58 \times 256$ \\
    12 & $1 \times 1$ convolution & $58 \times 58 \times 128$ \\
    13 & $3 \times 3$ convolution & $56 \times 56 \times 256$ \\
    14 & $2 \times 2$ AveragePooling & $28 \times 28 \times 256$ \\
    15 & $3 \times 3$ convolution & $26 \times 26 \times 512$ \\
    16 & $1 \times 1$ convolution & $26 \times 26 \times 256$ \\
    17 & $3 \times 3$ convolution & $24 \times 24 \times 512$ \\
    18 & $2 \times 2$ AveragePooling & $12 \times 12 \times 512$ \\
    19 & $3 \times 3$ convolution & $10 \times 10 \times 512$ \\
    20 & $1 \times 1$ convolution & $10 \times 10 \times 256$ \\
    21 & $3 \times 3$ convolution & $8 \times 8 \times 512$ \\
    22 & $1 \times 1$ convolution & $8 \times 8 \times 256$ \\
    23 & $3 \times 3$ convolution & $6 \times 6 \times 512$ \\
    24 & GlobalAveragePooling & 1 $\times$ 1 $\times$ 512 \\
    25 & FullyConnected & 2 or 10 \\ \hline
  \end{tabular}
  \caption{Our CNN architecture. The total number of trainable parameters is 8,328,608, except for the FullyConnected layer's parameters.}
  \label{tb:CNN_architecture}
\end{table}

In this section, we describe our CNN scheme. In our CNN, we apply a convolution of size $3\times 3$ kernels for deep multiple layers to extract characteristics over various scales. It is known that applying small sized kernels for multiple times is not only more computationally efficient but also enables more complex expressions compared to the network with a large sized kernel in single layer, e.g., when we convolve the  $3\times 3$ convolution kernel with five layers, we can refer to the pixels probed by an $11\times 11$ kernel in a single layer. The number of computation is simply $3\times 3\times 5$ for multiple small kernels and $11\times 11$ for single large kernel; therefore, multiple small kernels is more than twice as efficient.

We use the publicly available CNN platform Keras \citep{chollet2015keras} via a TensorFlow \citep{tensorflow2015-whitepaper} backend to construct our CNN. We follow the previous work \cite{Ribli2019a} for the architecture of the neural network, as summarized in Table \ref{tb:CNN_architecture}. The total number of training parameters in this architecture is $\sim 8\times 10^6$; therefore, $10^5$ data are required to avoid both over- and underfitting of the data \cite{Han2015}. Therefore, $6\times 10^5$ data should suffice.
 
We change the number of layers when we train our model. The reference number of layers and convolution sizes are summarized in Table \ref{tb:CNN_architecture}. If we halve the number of layers, the training and validation losses converge at a value ten times larger than the reference case and the validation accuracy is around 0.5, which means nothing for the classification. This is because this model is too simple. Conversely, if we double the number of layers, the losses do not decrease at all. This is because the number of trainable parameters is too large than the size of our training dataset and the vanishing gradients may occur \cite{2015arXiv151203385H}. Again, we observe that the validation accuracy fluctuates around 0.5.

In the $a \times b$ convolution layer, a feature map is generated from the $A \times B$ input image using the $a \times b$ kernels, which extract features from the input image. We set the stride of the convolution in our CNN to $1 \times 1$; then, the pixel value  ($F_{ij}$) at a position $i,j$ ($1 \le i \le A-(a-1)$, $1 \le j \le B-(b-1)$) in the feature map is
\begin{equation}
  \label{eq:conv}
    F_{ij} = \sum^{a-1}_{k=0} \sum^{b-1}_{l=0} I_{i+k,j+l} \times K_{k+1,l+1} \ ,
\end{equation}
where $I_{m,n}$ is the pixel value of the input image at position $m,n(1 \le m \le A, 1 \le n \le B)$ and $K_{m,n}$ is the pixel value of the kernel at $m,n (1 \le m \le a,1 \le n \le b)$. The value in the kernel is a weight parameter optimized  by training. Then, the feature map becomes the input image of the next layer.

After each convolution layer, we add a batch-normalization layer to normalize the distribution of the input feature map, increasing the training efficiency \citep{2015arXiv150203167I}.

In the $a \times b$ AveragePooling layer, if we set the stride to be the same as the pooling size, then $A \times B$ input image is converted into an $(A / a) \times (B / b)$ image. The pixel value ($F_{ij}$) at position $i,j(1 \le i \le A / a,\ 1 \le j \le B / b)$ in the output image is
\begin{equation}
  \label{eq:pooling}
  F_{ij} = \frac{1}{a \times b} \sum^{a-1}_{k=0} \sum^{b-1}_{l=0} I_{(A/a) \times i - k,(B/b) \times j - l} \ .
\end{equation}
In the AveragePooling layers, information in the input image is compressed and simplified.
In the GlobalAveragePooling layer, all pixel values in each input channel are averaged. The GlobalAveragePooling layer shows better performance and efficiency than the multiple FullyConnected layers \cite{Lin2013}. In a FullyConnected layer, the features extracted by the convolution layers are weighed by trainable weighting parameters, and the outputs from this layer are converted using the softmax activation function.

Activation function is introduced to make the mapping non-linear and thus makes the model more general and applicable for the complex dataset.
In our model,
ReLU is used after each convolutional layer and softmax is used after the FullyConnected layer; 
\begin{eqnarray}
  {\rm ReLU}(x) = \begin{cases}
  x & (x \ge 0) \\
  0 & (x < 0) \ ,
  \end{cases}
\end{eqnarray}
and 
\begin{equation}
  {\rm softmax}(x) = \frac{\exp(x)}{\displaystyle\sum_{k} \exp(x_k)},
\end{equation}
where the sum is taken over all nodes of FullyConnected layer so that the softmax function holds the condition of the probability.

Now, we can express the outputs in relation to the input and predicted classes,
\begin{equation}
  \label{eq:CNNout}
  \tilde{\boldsymbol{y}}(i|{\rm M}) =
  \{
  p_{1}(i|{\rm M}),
  p_{2}(i|{\rm M}),
  \cdots,
  p_{N}(i|{\rm M})
  \},
\end{equation}
where $p_{k}(i|{\rm M})$ is the probability that the CNN predicts the image as model $k$ given that the image is, in practice, taken from model M.

For optimization, we adopt a typical cross-entropy with the L2 regularization \citep{Ribli2019a} 
\begin{equation}
  \label{eq:loss}
  E(w) = - \sum_{k} y_{k} \ln{(\tilde{y}_{k})} + \lambda \sum_{\mathrm{kernel}} \sum_{i,j} K_{i,j}^2.
\end{equation}
In the first term,
the ground truth $y_k$ is the value of the $k$-th output class that takes one for the correct class, and zero otherwise. Prediction $\tilde{y}_k$ can take continuous values between 0 and 1. The second term is L2 regularization, which prevents our CNN from overfitting to the training data \citep{L2reg}. The value of $\lambda$ is determined on a the trial-and-error basis and we take $\lambda = 5 \times 10^{-5}$. This constant $\lambda$ determines the relative importance of this regularization. We see that the validation accuracy is better than in the case of $\lambda = 0$; for example, the validation accuracy increase by about 5\% in binary classification between CDM and 10 keV NCDM for random Gaussian simulation.

For optimization purposes, we use stochastic gradient descent. With the constant learning rate $\eta$, the weights $K_{i,j}$ and parameters in the FullyConnected layer are 
updated during the training by
\begin{equation}
    \label{eq:loss}
  w \rightarrow w - \eta \frac{\partial \bar{E}}{\partial w}.
\end{equation}
At the beginning of training, we set $\eta = 0.001$. Then, we multiply $\eta$ by $0.1$ at every 5 epochs after 10 (i.e. $10, 15, 20, 25$).
The weight updates are computed based on the averaged value of the loss function $\bar{E}$ over the mini-batch sample, which is randomly drawn from the training set. Here, we take 8 as the mini-batch sample size.
The choice of this mini-batch size is optimal in that, if we take the smaller sample size, it takes longer time to converge; however it cannot be larger due to limited memory resources.
After training, the value of $\bar{E}$ converges to between $\mathcal{O}(0.01)$ and $\mathcal{O}(0.1)$ depending on the NCDM mass.

We randomly keep 10\% of the training set untouched and use it for validation. The validation sample is not used to update the weight, but rather to monitor the validation loss at every training epoch. We stop the training if 
the validation loss averaged over the last 5 epochs converges to 1\%. 
For training binary classifications, the number of epochs it takes to converge is 50 for massive NCDM (e.g., $m_{\chi} \sim 20$ keV) but about 20 for less massive NCDM ($m_\chi\sim0.2$ keV).

\section{\label{sec:result} Results}

In this section, we compare the DM-model classification performance of our machine learning method with that using traditional correlation functions. In Section \ref{ssec:AUC}, we introduce a metric for quantifying the performances for CNN binary or multiple classification.
In Section \ref{ssec:result_class}, we show the result for binary
classification in discriminating between the 
CDM model and NCDM model with some specific masses. Finally, in Section \ref{ssec:result_regression}, we show multi-class classification.
For later convenience, we denote the different model discrimination schemes as follows: CNN-NS, CNN on N-body simulations; CNN-RG, CNN on random Gaussian simulations; and TPCF, two-point correlation function.

\subsection{\label{ssec:AUC} Evaluation of CNN}
The area under the precision-recall (PR) curve
is used to quantify the performance of CNN in our work.
Our method predicts the probability that an image will be generated from the model. Therefore, we can decide whether the image corresponds to the given model. We do this by introducing the variable threshold $t$. If $p_{k}(i)>t$, we recognize that the $i$-th image is classified as model $k$. Therefore, we can consider four different cases for this type of game, given that we focus on the $k$-th NCDM model:
\begin{itemize}
    \item True Positive (TP): $p_k(i|k)>t$; the image of the $k$-th NCDM model is correctly classified as model $k$.
    \item True Negative (TN): $p_k(i|j)<t$; the image of model $j \neq k$ is correctly not classified as the $k$-th NCDM model.
    \item False Positive (FP): $p_k(i|j)>t$; the image of model $j \neq k$ is mis-classified as the $k$-th NCDM model.
    \item False Negative (FN): $p_k(i|k)<t$; the image of the $k$-th NCDM model is not classified as the $k$-th NCDM model.
\end{itemize}
Now, the PR curve can be defined as the collection of points at which parameter $t$ continuously changes from 0 to 1.
\begin{equation}
  \mathrm{PR}:
  x(t)=\frac{\rm TP}{\rm TP+FN},
  y(t)=\frac{\rm TP}{\rm TP+FP},
\end{equation}
where $x$ and $y$ are sometimes recognized as recall (sensitivity) and precision.
the area under the curve (AUC) is in the range of $a_{\mathrm{rand}}$ to $1$; it takes values close to unity when the classifier can distinguish the models efficiently and close to 0.5 for the binary classification or 0.1 for the ten(multi)-class classification if the classifier does nothing about the model discrimination.\\


\subsection{\label{ssec:result_class}Binary Classification}
\begin{figure*}[tp]
\centering
\begin{tabular}{cc}
  \includegraphics[keepaspectratio,width=0.5\linewidth]{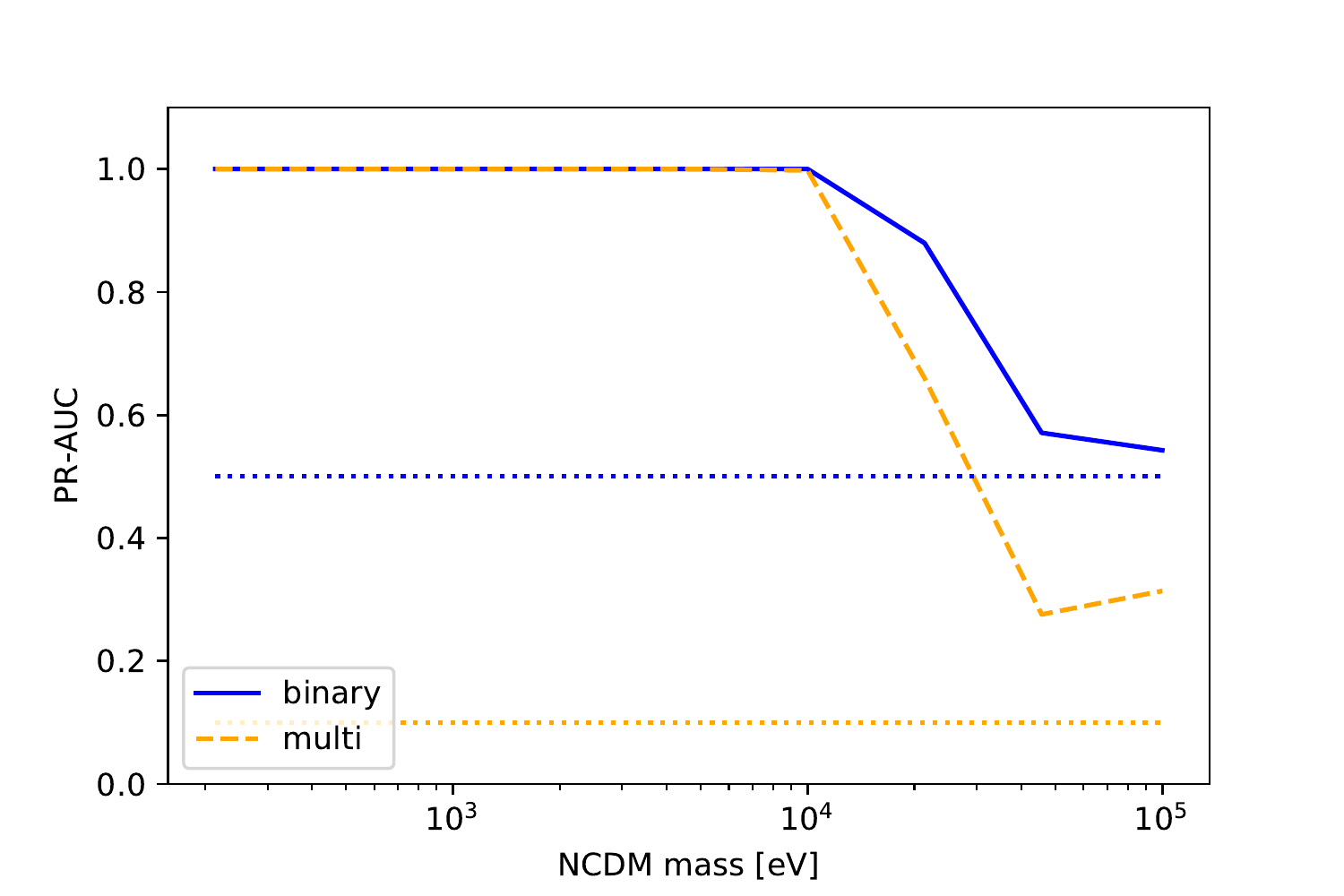} &
  \includegraphics[keepaspectratio,width=0.5\linewidth]{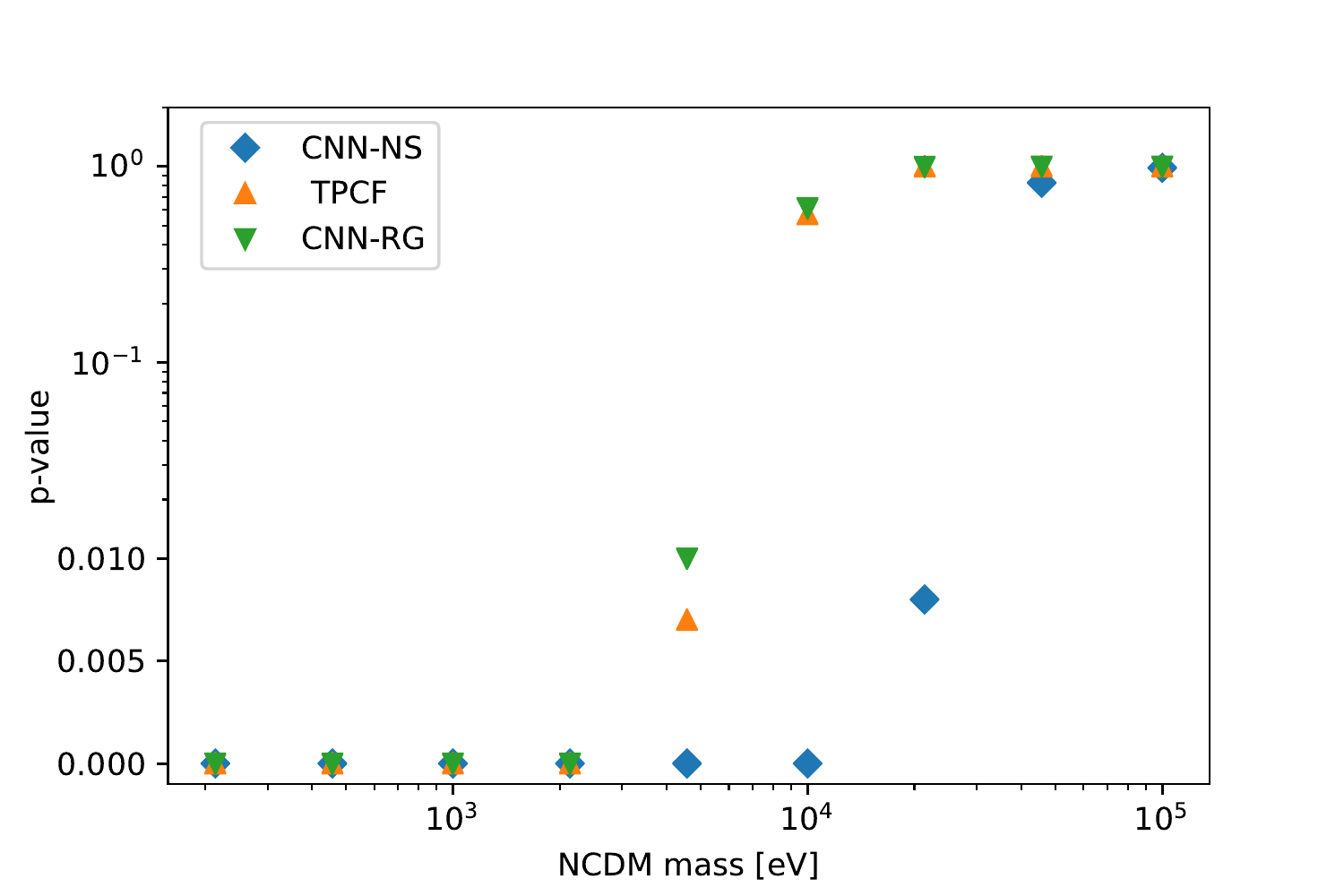}
\end{tabular}
  \caption{(Left)
  AUC as a function of the NCDM mass for binary (blue solid) and multiple (orange dashed) classifications. Dashed horizontal lines represent AUC values when the CNN cannot discriminate the models at all (i.e., random classification).
  (Right)
  Comparison of model discriminations probed by $p$-value for TPCF, CNN, and CNN-RG with binary classification. We see that the CNN-RG's p-value is almost consistent with that from TPCF, and that the CNN has a significantly better p-value than fir the other two schemes.
  }
  \label{fig:p-value}
\end{figure*}

First, we show the performance of CNN-NS for the binary
classification 
between the CDM and NCDM models using AUC.
Fig. \ref{fig:p-value} shows the AUC curve as a function of the mass of NCDM.
The AUC is greater than $0.99$ when our CNN distinguishes CDM from NCDM $m_{\chi} \le 1$ keV. Moreover, we can distinguish CDM from $21$ keV NCDM with an accuracy of AUC $= 0.88$.
With our current simulation resolution, it is difficult to discriminate the NCDM models with masses of $46$ or $100$ keV from the CDM model, for which the AUCs are 0.57 and 0.54, respectively (close to the limit of random classification).

Next, we compare the performance of classification by CNN-NS, TPCF, and CNN-RG using the $p$-value to show that CNN-NS can harness more information and offer better performance than TPCF.
We consider the null hypothesis that the simulated data from the NCDM model do not differ from those from the CDM model. The $p$-value can be calculated as the probability of having a larger $\chi^2$ value than
\begin{equation}
    \chi^2_k = \mbf{D}(k) \textbf{C}^{-1} \mbf{D}(k),
\end{equation}
where for CNN,
\begin{equation}
    \mbf{D}(k) \equiv \frac{1}{N}
    \sum^{N}_{i=1} [
    \tilde{\mbf{y}}(i|k) - \tilde{\mbf{y}}(i| \mathrm{CDM})],
\end{equation}
for TPCF,
\begin{equation}
    \mbf{D} \equiv w_P (r|k) - w_P(r|\mathrm{CDM})
\end{equation}
and the argument $k$ denotes the model index to be tested.
The covariance matrix for TPCF has already been given by Eq. (\ref{eq:jk_cov}). For CNN, it can be defined about the CDM model as
\begin{equation}
    \begin{split}
      \mathrm{C}_{jk} =
      \frac{1}{N} \displaystyle\sum^{N}_{i} &[ (p_{j}(i | \mathrm{CDM}) - \langle p_{j}(\mathrm{CDM}) \rangle) \\
      &\times (p_{k}(i | \mathrm{CDM}) - \langle p_{k}(\mathrm{CDM}) \rangle)
    \end{split}
\end{equation}
where $j$ and $k$ represent the indices of the model, $\langle p_{j}(\mathrm{CDM}) \rangle = \frac{1}{N} \displaystyle\sum^{N}_{i} p_{j}(i | \mathrm{CDM})$, and $N$ is the number of test images for CDM simulation.

The right-hand panel of Fig. \ref{fig:p-value} shows the $p$-value which represents the probability of the input data being consistent with the CDM model. We see that the CNN-NS shows the best performance among the three methods. For example, with the CNN-RG and TPCF, the CDM model is rejected only less than 3-$\sigma$ for $m_\chi=4.6$ keV; however, the $p$-value for CNN-NS is limited by the numerical precision and we can recognize it as absolutely zero. For CNN-NS, $m_\chi=21 $keV can be rejected more than 2-$\sigma$.

\subsection{\label{ssec:result_regression}Multiple Class Classification}
We also perform multiple class classification for ten DM models (9 NCDM $+$ CDM) to show the CNN's potential to constrain the particle mass.

First, we calculate the AUC and show the result in
Fig. \ref{fig:p-value}. AUCs are greater than $0.99$ for
$m_\chi \le 10$ [keV], and
CNN-NS can identify
these NCDM models.
However, for more massive NCDM models,
the classification becomes
more difficult. For the $21$ keV NCDM, AUC is $0.66$, and for the $41$ and $100$ keV NCDM models, the AUCs are less than $0.3$ close to the limit of random classification.

Now, let us consider $\langle p(M) \rangle$ for the input data of model, which is defined as
\begin{equation}
  \langle p_{M} \rangle = \frac{1}{N }\displaystyle\sum^N_{i} p_M(i|M)
\end{equation}
where $N$ is the number of test images of model M. $\langle p_{M} \rangle$ represents the probability that the test data are taken for model M. When our CNN can distinguish the model M data from the other models, $\langle p_{M} \rangle$ is close to unity. However, $\langle p_{M} \rangle$ is close to 0.1 if our CNN cannot discriminate the models at all. $\langle p_{M} \rangle$ can be calculated for data for which we cannot know the correct label (e.g., observational data) in future work.
Our CNN can obtain correct results almost perfectly for the NCDM model with $m_\chi<10$ keV, 
and $\langle p_{M} \rangle \ge 0.99$.
In addition, our CNN offers high performance for the $21$ keV NCDM model with  
$\langle p_{M} \rangle$ = 0.89, whereas
the performances for the CDM and NCDM model with $m_\chi \ge 46$ keV  are low,
with $\langle p_{M} \rangle < 0.3$.

\section{\label{sec:discussion} Discussion}
First, we compare our $p$-value results to those from random Gaussian simulations to shed light on the additional information attained from machine learning.

A random Gaussian simulation is a particular realization of a density field with an expected power spectrum and for which the phase of the density field variable is totally random with no correlations. 

We do not see any higher-order (e.g., bispectrum or trispectrum) correlations under the random Gaussian simulation.  Therefore, given that the statistical properties fully determine the power of the constraints upon the parameters, we expect the CNN-RG to have comparable power constraints to those of the TPCF. The right-hand panel of Fig. \ref{fig:p-value} shows that the constraint abilities of the CNN-RG and TPCF schemes are almost identical. Therefore, we can conclude that the CNN can extract all statistical information in the case of random Gaussian simulation and use it to constrain the underlying parameters. We would like to stress that, although the ML method is usually considered to be a black box, it never adds extra information that is not potentially included in the data.

On the other hand, CNN-NS offers
better performance than the  
other two methods. This is because the distribution of dark matter particles in the N-body simulation contains more 
information than the TPCF
and the CNN-NS 
has an ability to extract them.
Therefore, we conclude that the CNN is useful for constraining the cosmological model through the dark matter distribution. 

We note that the analysis in this paper assumes that DM and luminous objects such as galaxies are distributed in the same manner; in practice, this is not true. One solution to relax this unrealistic assumption is to use cosmic shear fields \citep{2020arXiv200706529Z}. The authers of \cite{2020arXiv200706529Z} estimated the expected constraints on the matter density today, $\Omega_m$, and the amplitude of the matter power spectrum, $\sigma_8$, using CNN on the convergence field under weak lensing observation. The result is compared with other statistics such as the power spectrum peak counts and the Minkowski functionals; it is found that the CNN best constrains the cosmological parameters of interest. To apply our analysis on the real photometric observation data, we must see how the light distribution from galaxies is related to that of DM; however, this is beyond the scope of this paper and we will revisit it in our future work.

Next, let us show the difference between binary and multiple classifications.
In both cases, the AUC is almost unity for distinguishing the $m_\chi<10$keV NCDM model, relatively worse at $m_\chi=21$keV, and CNN-NS has almost no power of discrimination for $m_\chi>46$keV model (i.e., it approaches the dotted lines). Therefore, the results obtained here are independent of the type of classification.

In multiple classification, $\langle p_{M} \rangle$ is greater than 0.99 for the NCDM model $m_\chi \le 10$ keV and is 0.89 for $m_\chi = 21$ keV. Thus, our CNN offers good performance in identifying the NCDM models $m_\chi \le 21$ keV for the N-body simulation data. In our future work, we need more realistic training or test datasets (e.g., hydrodynamic simulation) and a method for evaluating our CNN's results. This must be done before it can be applied to observation data.

\section{\label{sec:summary}Summary}
In this paper, we have shown that the CNN can distinguish DM masses for the NCDM model better than conventional correlation function analysis. Moreover, we have shown that the CNN can fully extract the statistical information contained in a random Gaussian simulation and that can be extracted by the source of information content that the CNN originates from the non-linear gravitational evolution of the large-scale structure.
To see this, 
we  perform a suite of N-body simulations with different DM particle masses.
The simulation data are projected along the line of sight assuming a photometric galaxy survey. In our analysis, we assume that the line-of-sight resolution is 50 Mpc/$h$, and we observe the unbiased tracer of DM.

In binary classification, we compare the images of the large-scale structure for CDM with the NCDM models with different masses.
The results are compared using a conventional two-point correlation function. In addition, we repeat our CNN analysis for the random Gaussian simulation to see if the 
CNN extracts and entirely exhausts the statistical information. 
Hence, we find that, CNN-NS offers a better performance than TPCF. The TPCF cannot distinguish the CDM model from the NCDM models with $m_\chi \ge 10$ keV,
whereas CNN-NS can distinguish the CDM model from the NCDM models with $m_\chi < 21$ keV.
If we compare the p-values of the CNN-NS and TPCF schemes for distinguishing the NCDM and CDM models, we see that CNN-NS offers superior performance. In addition, the $p$-values from CNN-RG and TPCF are almost identical.
Therefore, CNN uses all statistical information 
contained in the random Gaussian simulations.

In the multiple classification scheme, we investigate how well the DM model can be distinguished from the other models.  
Our CNN shows
good performance in terms of $\langle p_M \rangle$ for $m_\chi \le 21$ keV.  
Therefore, the CNN model
can identify these NCDM models with high accuracy in the N-body simulation data when we know the distribution of DM particles.

We also compare binary to multiple classification using the area under the precision-recall curve. Both classification offer the similar performance in terms of the DM mass, which can be discriminated by CNN. 
Therefore, our results do not depend on the type of classification.

Our work shows the potential for CNN to  constrain the DM mass more strongly by analyzing the large-scale structure of the universe.
However, we have thus far only used CNN classification on simulation data for which the DM distribution is known.
In practice, we cannot see the true underlying DM
directly, but can only trace it from observing galaxies.
It is important to extend our analysis to  practical observables such as the light distribution from luminous objects, or unresolved background radiation. We will return to this topic in our future work.

\section*{acknowledgment}
We are grateful to Kiyotomo Ichiki, Hironao Miyatake and Shiro Ikeda for fruitful discussions. This work is supported by Japan Science and Technology Agency (JST) AIP Acceleration Research Grant Number JP20317829 and JSPS Kakenhi Grant number JP18H04350. Part of the computation is performed on Cray xc50 and GPU cluster at CfCA in NAOJ and GPU workstation at Nagoya University.

\bibliography{bibtex}

\begin{thebibliography}{10}

\bibitem{Boyarsky2019}
A.~{Boyarsky}, M.~{Drewes}, T.~{Lasserre}, S.~{Mertens}, and O.~{Ruchayskiy},
\newblock Progress in Particle and Nuclear Physics {\bf 104}, 1 (2019),
  1807.07938.

\bibitem{Alvarez2020}
A.~{Alvarez} {\em et~al.},
\newblock arXiv e-prints , arXiv:2002.01229 (2020), 2002.01229.

\bibitem{Garzilli2019}
A.~{Garzilli}, O.~{Ruchayskiy}, A.~{Magalich}, and A.~{Boyarsky},
\newblock arXiv e-prints , arXiv:1912.09397 (2019), 1912.09397.

\bibitem{Garzilli2019a}
A.~{Garzilli} {\em et~al.},
\newblock \mnras {\bf 489}, 3456 (2019), 1809.06585.

\bibitem{Ribli2019a}
D.~{Ribli} {\em et~al.},
\newblock \mnras {\bf 490}, 1843 (2019), 1902.03663.

\bibitem{Ribli2019}
D.~{Ribli}, B.~{\'A}. {Pataki}, and I.~{Csabai},
\newblock Nature Astronomy {\bf 3}, 93 (2019), 1806.05995.

\bibitem{Pan2019}
S.~{Pan} {\em et~al.},
\newblock arXiv e-prints , arXiv:1908.10590 (2019), 1908.10590.

\bibitem{Bonjean2020}
V.~{Bonjean},
\newblock \aap {\bf 634}, A81 (2020), 1911.10778.

\bibitem{Peel2019}
A.~{Peel} {\em et~al.},
\newblock \prd {\bf 100}, 023508 (2019), 1810.11030.

\bibitem{Modi2018}
C.~{Modi}, Y.~{Feng}, and U.~{Seljak},
\newblock \jcap {\bf 2018}, 028 (2018), 1805.02247.

\bibitem{Akrami2018}
{Planck Collaboration} {\em et~al.},
\newblock arXiv e-prints , arXiv:1807.06209 (2018), 1807.06209.

\bibitem{Boyanovsky2011}
D.~{Boyanovsky} and J.~{Wu},
\newblock \prd {\bf 83}, 043524 (2011), 1008.0992.

\bibitem{Lesgourgues2011a}
J.~{Lesgourgues},
\newblock arXiv e-prints , arXiv:1104.2932 (2011), 1104.2932.

\bibitem{Ma1995}
C.-P. {Ma} and E.~{Bertschinger},
\newblock \apj {\bf 455}, 7 (1995), astro-ph/9506072.

\bibitem{Dodelson1994}
S.~{Dodelson} and L.~M. {Widrow},
\newblock \prl {\bf 72}, 17 (1994), hep-ph/9303287.

\bibitem{Lesgourgues2011b}
J.~{Lesgourgues} and T.~{Tram},
\newblock \jcap {\bf 2011}, 032 (2011), 1104.2935.

\bibitem{Crocce2006}
M.~{Crocce}, S.~{Pueblas}, and R.~{Scoccimarro},
\newblock \mnras {\bf 373}, 369 (2006), astro-ph/0606505.

\bibitem{Springel2005}
V.~{Springel},
\newblock \mnras {\bf 364}, 1105 (2005), astro-ph/0505010.

\bibitem{Euclid}
R.~{Laureijs} {\em et~al.},
\newblock arXiv e-prints , arXiv:1110.3193 (2011), 1110.3193.

\bibitem{LSST}
{\v Z}.~{Ivezi{\'c}} {\em et~al.},
\newblock \apj {\bf 873}, 111 (2019), 0805.2366.

\bibitem{Lupton+:1999}
R.~H. {Lupton}, J.~E. {Gunn}, and A.~S. {Szalay},
\newblock \aj {\bf 118}, 1406 (1999), astro-ph/9903081.

\bibitem{2015ascl.soft08007J}
M.~{Jarvis},
\newblock {TreeCorr: Two-point correlation functions}, 2015, 1508.007.

\bibitem{chollet2015keras}
F.~Chollet {\em et~al.},
\newblock Keras,
\newblock \url{https://keras.io}, 2015.

\bibitem{tensorflow2015-whitepaper}
M.~Abadi {\em et~al.},
\newblock {TensorFlow}: Large-scale machine learning on heterogeneous systems,
  2015,
\newblock Software available from tensorflow.org.

\bibitem{Han2015}
S.~{Han}, J.~{Pool}, J.~{Tran}, and W.~J. {Dally},
\newblock arXiv e-prints , arXiv:1506.02626 (2015), 1506.02626.

\bibitem{2015arXiv151203385H}
K.~{He}, X.~{Zhang}, S.~{Ren}, and J.~{Sun},
\newblock arXiv e-prints , arXiv:1512.03385 (2015), 1512.03385.

\bibitem{2015arXiv150203167I}
S.~{Ioffe} and C.~{Szegedy},
\newblock arXiv e-prints , arXiv:1502.03167 (2015), 1502.03167.

\bibitem{Lin2013}
M.~{Lin}, Q.~{Chen}, and S.~{Yan},
\newblock arXiv e-prints , arXiv:1312.4400 (2013), 1312.4400.

\bibitem{L2reg}
A.~E. Hoerl and R.~W. Kennard,
\newblock Technometrics {\bf 12}, 55 (1970).

\bibitem{2020arXiv200706529Z}
J.~M. {Zorrilla Matilla}, M.~{Sharma}, D.~{Hsu}, and Z.~{Haiman},
\newblock arXiv e-prints , arXiv:2007.06529 (2020), 2007.06529.

\end{thebibliography}
\bibliographystyle{h-physrev}

\end{document}